\documentclass{article}
\usepackage{icmc2023template}
\usepackage{times}
\usepackage{ifpdf}
\usepackage{soul}
\usepackage{url}
\usepackage{caption}
\usepackage[english]{babel}
\def\papertitle{LaunchpadGPT: Language Model as Music Visualization\\ Designer on Launchpad}
\def\firstauthor{Siting Xu$^{*}$}
\def\secondauthor{Yunlong Tang$^{*}$}
\def\thirdauthor{Feng Zheng$^{\dagger}$}

\newif\ifpdf
\ifx\pdfoutput\relax
\else
   \ifcase\pdfoutput
      \pdffalse
   \else
      \pdftrue
  \fi
\fi

\ifpdf % compiling with pdflatex
  \usepackage[pdftex,
    pdftitle={\papertitle},
    pdfauthor={\firstauthor, \secondauthor, \thirdauthor},
    bookmarksnumbered, % use section numbers with bookmarks
    pdfstartview=XYZ % start with zoom=100% instead of full screen; 
   ]{hyperref}
  \usepackage[pdftex]{graphicx}
  \graphicspath{{./figures/}}
  \DeclareGraphicsExtensions{.pdf,.jpeg,.png}

  \usepackage[figure,table]{hypcap}

\else % compiling with latex
  \usepackage[dvips,
    bookmarksnumbered, % use section numbers with bookmarks
    pdfstartview=XYZ % start with zoom=100% instead of full screen
  ]{hyperref}  % hyperrefs are active in the pdf file after conversion

  \usepackage[dvips]{epsfig,graphicx}
  \graphicspath{{./figures/}}
  \DeclareGraphicsExtensions{.eps}

  \usepackage[figure,table]{hypcap}
\fi

\hypersetup{
    colorlinks,%
    citecolor=black,%
    filecolor=black,%
    linkcolor=black,%
    urlcolor=black
}

\title{\papertitle}

\oneauthor
   {\firstauthor,~\secondauthor,~\thirdauthor}{Southern University of Science and Technology \\ 
     {\tt \{\href{mailto:xust2019@mail.sustech.edu.cn}{xust2019}, 
           \href{mailto:tangyl2019@mail.sustech.edu.cn}{tangyl2019}\}@mail.sustech.edu.cn, 
           \href{mailto:zhengf@sustech.edu.cn}{zhengf@sustech.edu.cn}}}

\begin{document}
\capstartfalse
\maketitle
\capstarttrue
{\let\thefootnote\relax\footnotetext{$^*$ Equal contribution. $^{\dagger}$ Corresponding author.}}
%最后再改
\begin{abstract}
Launchpad is a musical instrument that allows users to create and perform music by pressing illuminated buttons. To assist and inspire the design of the Launchpad light effect, and provide a more accessible approach for beginners to create music visualization with this instrument, we proposed LaunchpadGPT model to generate music visualization designs on Launchpad automatically. Based on the language model with excellent generation ability, our proposed LaunchpadGPT takes an audio piece of music as input and outputs the lighting effects of Launchpad-playing in the form of a video (Launchpad-playing video). We collect Launchpad-playing videos and process them to obtain music and corresponding video frame of Launchpad-playing as prompt-completion pairs, to train the language model. The experiment result shows the proposed method can create better music visualization than random generation methods and hold the potential for a broader range of music visualization applications. Our code is available at \href{https://github.com/yunlong10/LaunchpadGPT}{\textcolor{black}{https://github.com/yunlong10/LaunchpadGPT}}.
\end{abstract}

\section{Introduction}\label{sec:introduction}

\begin{figure}[h]
\centering
\includegraphics[width=0.75\columnwidth]{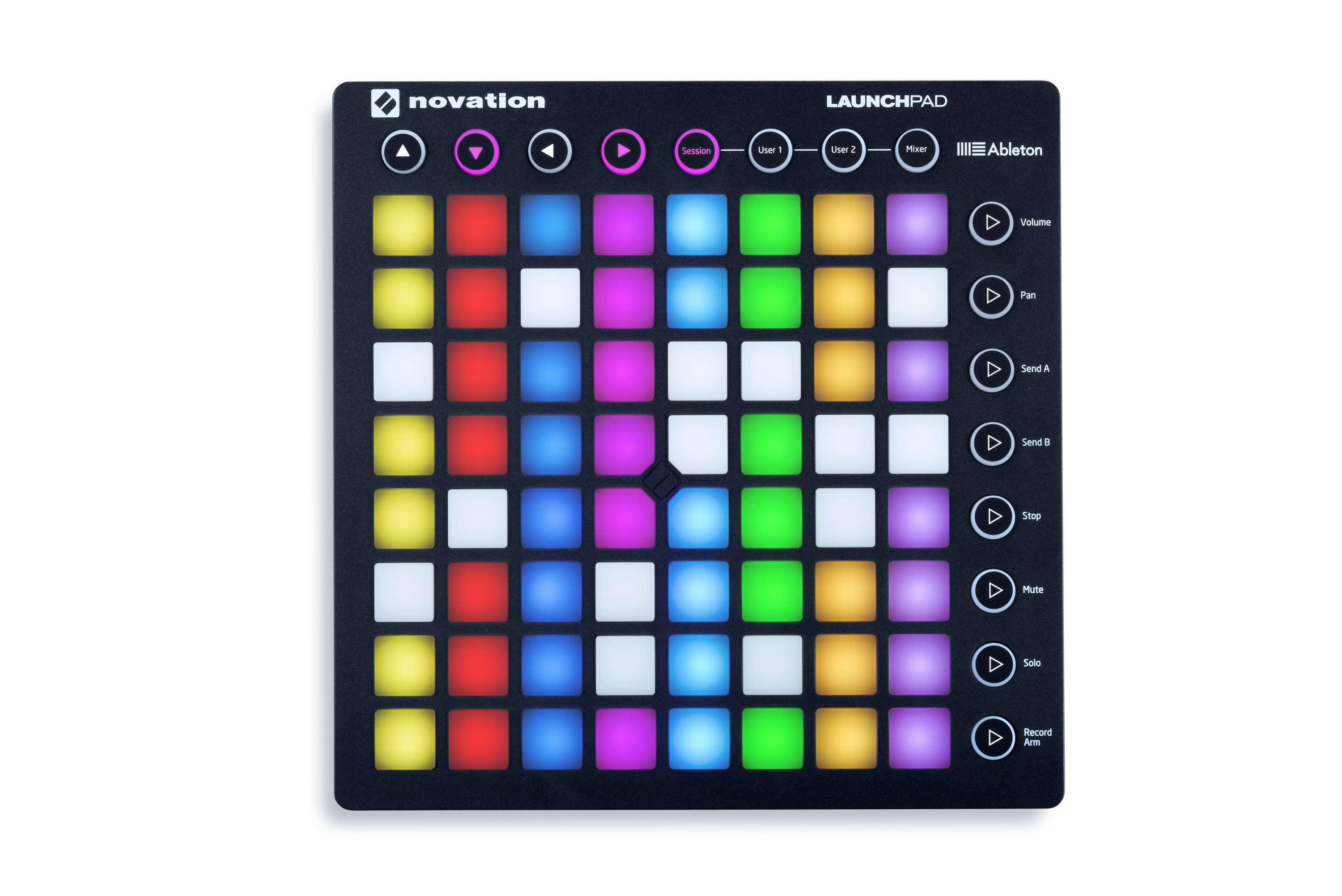}
\caption{This is an illustrative figure of a Launchpad with 8×8 illuminated buttons from Novation.}
\label{fig:launchpad}
\end{figure}

Launchpad is a popular musical hardware controller that allows users to create and perform music by pressing illuminated buttons or programming. It consists of 8×8 illuminated buttons, allowing users to trigger various musical elements like sounds, samples, and loops. Launchpad offers operations for music production and live performances. %介绍Launchpad是什么
Due to the ornamental nature of Launchpad, enthusiasts frequently share their performance videos online, providing audiences with immersive visual experiences. %Launchpad的视觉特性
Figure \ref{fig:launchpad} illustrates a Launchpad\footnote{https://www.ableton.com/en/products/controllers/launchpad/}. 

By leveraging its illuminated buttons, Launchpad can generate synchronized visual effects that enrich the experience for both performers and audiences, which is an excellent form of music visualization. These visual effects can be programmed and customized to match the music's mood and style, resulting in audiovisual performances that fuse sight and sound \cite{lima2021survey}. %音乐可视化

% Our work focuses on the visual attributes of Launchpad. 
Building Launchpad projects can be a complex and time-consuming task, particularly for beginners, for it involves multiple steps such as music production, programming, and designing intricate light effects. 
Besides, the design of light effects for Launchpad projects shares common ground with the broader field of music visualization, as it requires careful consideration of factors such as the rhythm and tempo of the music, the visual impact desired, and the synchronization with the audio. 
% Each of these steps requires time and expertise, especially the process of light design. 

% Light effects design provides an xxx possibility to extend the visual effects of Launchpad to more general music visualization. 
% It involves experimenting with different lighting patterns, color schemes, and dynamic effects to create immersive experiences.

To address these issues, we formulate the Launchpad visualization as translating a piece of music to a music visualization generation problem.
Since language models show an excellent ability in cross-modal processing and the visual effects of music visualization can be encoded as similar semantic features as text information, we adopted the Generative Pre-trained Transformer (GPT) \cite{gpt1,gpt2,gpt3} and propose a language model-based framework for Launchpad-playing video generation, named \textbf{LaunchpadGPT}. 
Specifically, we first build a dataset of music-frame pairs by processing Launchpad-playing videos from the web to allow LaunchpadGPT to train and automatically generate music visualization. 
We then convert the Launchpad-playing frames into a textual representation of RGB values and button coordinates noted as X, followed by LaunchpadGPT to automatically generate Launchpad-playing videos. 
For the training and inference phase, the Mel-Frequency Cepstral Coefficients (MFCC) features extracted from audio act as prompt, and the corresponding RGB values with X coordinates act as completion. 

% leverage language model to perform the music visualization on Launchpad. Recently, language models show an exploding ability in cross-modal processing. Besides, the visual effects of music visualization display a certain level of semantic information. To extend the cross-modal ability of language model to the music visualization domain, we adopted the Generative Pre-trained Transformer (GPT) \cite{gpt1,gpt2,gpt3} and train a new model, \textbf{LaunchpadGPT}, which holds the ability to automatically generate Launchpad-playing videos. We build a dataset of music-frame pairs by processing Launchpad-playing videos, which allows LaunchpadGPT to train and automatically generate music visualization. 

Our model extends its ability to the domain of music visualization and video generation, enabling it to learn the complex patterns and structures of Launchpad-playing. It simplifies light design by providing automatic schemes based on inputted music. Additionally, it enables the creation of synchronized music videos showcasing Launchpad's performance and dynamic lighting effects, reducing production time.
Experimental results demonstrate that LaunchpadGPT reveals the potential to extend its capabilities to other music visualization applications. Our envisioned applications contain a range of design possibilities within the realm of music visualization, including charting music game levels, LED screen designs for music performance venues, and devising captivating lighting schemes for music dance floors.

In summary, our contribution is twofold:
\begin{itemize}
    \item We proposed LaunchpadGPT, a language model based on GPT, to generate music visualization on Launchpad automatically with given music.
    \item We collected a dataset of Launchpad-playing videos and construct prompt-completion pairs for training, which bridges the music and vision with texts.
\end{itemize}

% \section{Preliminary}\label{sec:preliminary}
% \subsection{Launchpad}\label{subsec:launchpad}
% Launchpad is a hardware device designed to enhance music creation, performance, and live experience of music. It consists of an 8×8 grid of illuminated buttons, allowing users to trigger various musical elements like sounds, samples, and loops. Launchpad serves as a versatile tool for composers, producers, and musicians, providing control over software and hardware instruments, sequencers, and digital audio workstations (DAWs). Users can manipulate musical arrangements, trigger loops, and launch clips, and adjust volume levels through interaction with the Launchpad's buttons and controls. Launchpad is compatible with popular software such as Ableton Live, offering versatility for musicians and producers across different setups. Figure \ref{fig:launchpad} is an illustrative figure of Launchpad\footnote{https://www.ableton.com/en/products/controllers/launchpad/}.
%来点图展示一下

\section{Method}\label{sec:method}
\subsection{Overview}\label{subsec:overview}

\begin{figure*}[t]
  \centering
  \includegraphics[width=\textwidth]{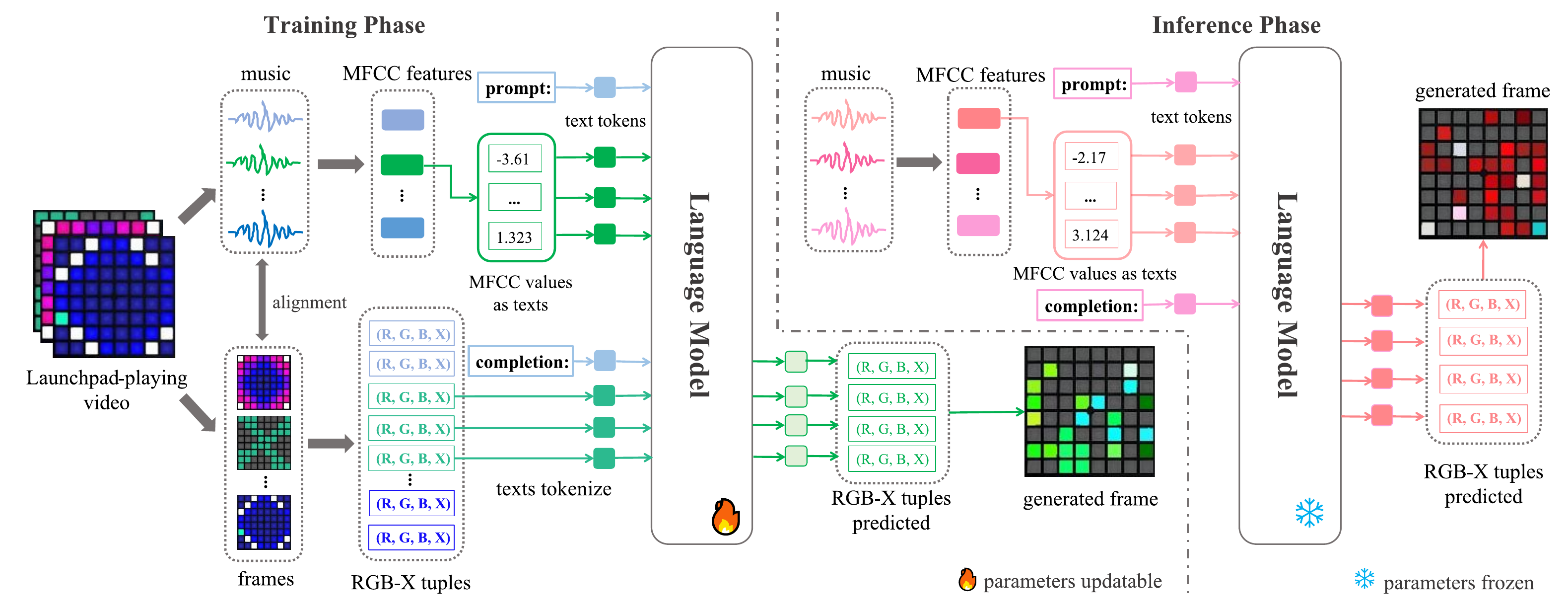}
  \caption{The framework of LaunchpadGPT. In the training phase, aligned music and frame information is extracted from the Launchpad-playing videos. MFCC features are extracted from the music, while the color-coordinate tuple (R, G, B, X) is obtained from the frames. The MFCC features to serve as the text prompt input for the language model, while RGB-X tuples are also tokenized as texts for ``completion'', training the language model in the teach-forcing paradigm. In the inference phase, the input music's features are extracted and transferred to text tokens, as prompt input for the trained language model, generating a series of (R, G, B, X) tuples as the ``completion'' output, and guiding the frame generation.}
  \label{fig:framework}
\end{figure*}

Figure \ref{fig:framework} shows the framework of our LaunchpadGPT. Given a Launchpad-playing video, the video frames and music (audio) will be extracted respectively. For music, Mel-Frequency Cepstral Coefficients (MFCC) features will be extracted, whose number will be the same as the number of video frames so that the audio can be aligned well with the frames. The buttons' color information of the Launchpad keyboard in the frame will be recorded in text form called an RGB-X tuple. The X is the coordinate of one button on the keyboard. Then, we take the MFCC features as texts and concatenate them with the corresponding frames' RGB-X tuples (also in text form) to get the prompt-completion pairs.

In the training phase, the prompt-completion pairs are the input of a language model, which will be trained in a teacher-forcing paradigm. In the inference phase, the prompts (MFCC texts) will be the input to predict the completion (RGB-X tuples). The generated RGB-X will be used to generate the frames of the Launchpad keyboard.

\subsection{Feature Extraction}\label{subsec:feature_extraction}

In order to align the features of the music and the video frames, we set the audio frame length $N_{af}$ and audio hop length $N_{ah}$ as following:
\begin{equation}
    N_{ah} = \frac{N_a}{N_{vf}-1},~N_{af} = 2\cdot N_{ah},
\label{eq:len}
\end{equation}

where $N_a$ is the audio sampling number, $N_{vf}$ is the number of video frames. Then we will get $N_{vf}$ MFCC features with the dimension of 128.
For video frames, we represent them with (R, G, B, X) tuples, where the R, G, and B are the red, green, and blue channels of each button, and the X, ranging from 0 to 63, is the coordinate of one button on the Launchpad keyboard. Thus, we will get 64 RGB-X tuples from each frame.

\subsection{Prompt-Completion Pairs Construction}

\begin{figure}[t]
\centering
\includegraphics[width=\columnwidth]{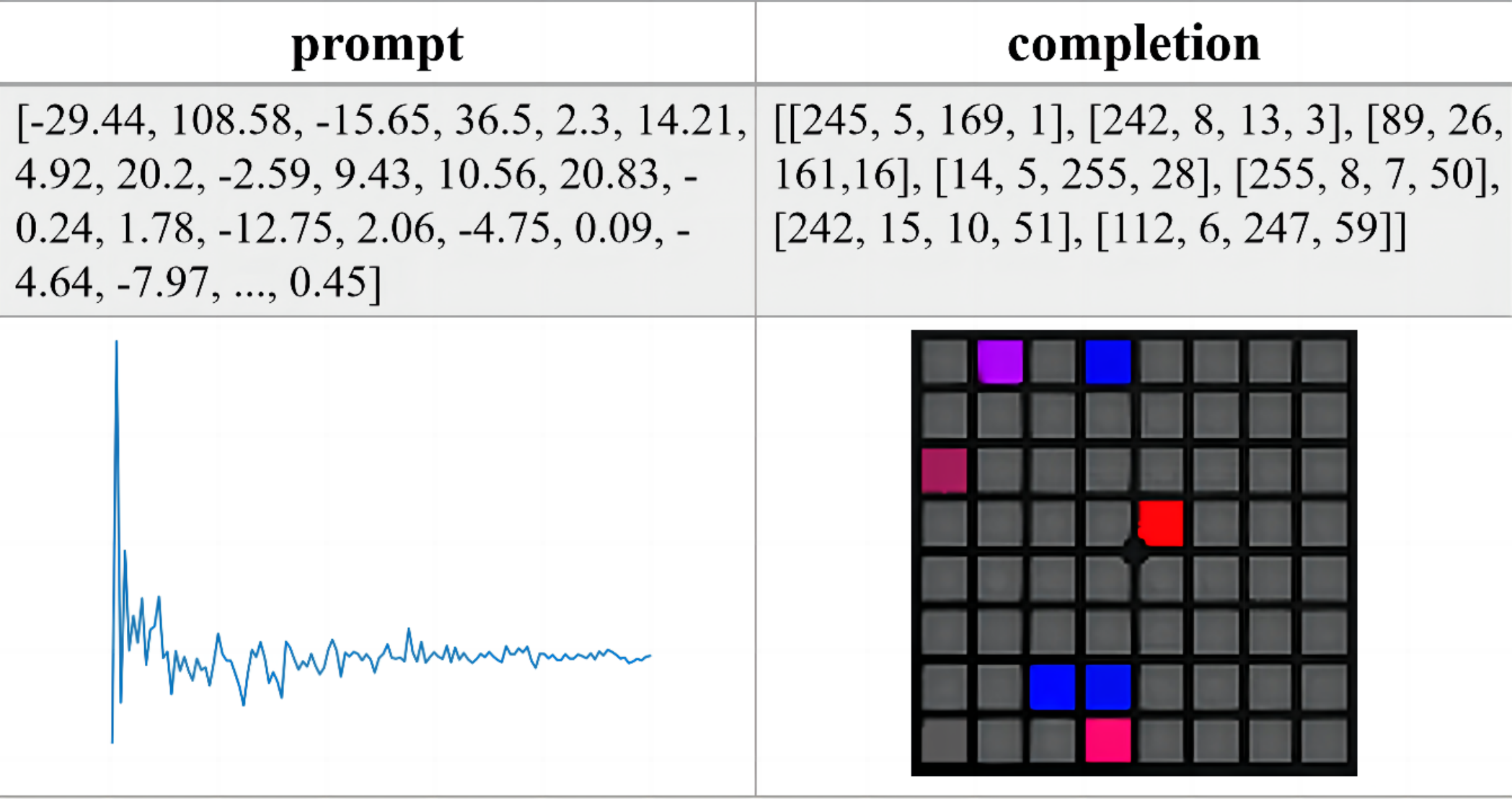}
\caption{An example of prompt-completion pair. After the ``prompt:'' are the textual MFCC feature values, and after the ``completion:'' are the textual RGB-X tuples. The first RGB-X tuple (245, 5, 169, 1) denotes that the second button (the index of the first button is 0) on the Launchpad keyboard is purple.}
 \label{fig:example}
\end{figure}

With the MFCC features of music and corresponding frames' color-coordinate information, we can construct the prompt-completion pairs for the training and inference of the language model. Specifically, the MFCC features are transferred to texts directly as prompts with an additional prefix ``prompt:''. The corresponding frame's tuples of RGB-X are also transferred to texts with the prefix ``completion:''. Figure \ref{fig:example} shows a prompt-completion pair example.

\subsection{NanoGPT}\label{subsec:nano_gpt}
A language model refers to a probability distribution over sequences of tokens. 
A sequence is taken as an input for a language model and a probability performed as an assessment act as output, which can be denoted as 
$p(x_{1:L})$, 
with $x_{1:L}$ as a character sequence sampled by the language model, and $p$ as the probability.
The calculation of $p$ is to use the chain rule of probability: 
\begin{equation}
    p(x_{1:L})=\prod_{i=1}^{L}p(x_i\mid x_{1:i - 1}).
\label{eq:LM}    
\end{equation}

Usually, conditional generation is to take a prefix sequence $x_{1:i}$ as input and sample the rest $x_{i+1:L}$ as output. The input is called \textbf{prompt} and the output is called \textbf{completion}. In our work, the MFCC features extracted from the music will be taken as prompt and a series of tuples representing the colors and coordinates of the buttons on Launchpad which is generated by LaunchpadGPT is the completion.
Therefore, given a prompt $x_{1:i}$, an autoregressive language model will generate a completion $x_{i+1:L}$ accordingly.

The language model we use is NanoGPT\footnote{https://github.com/karpathy/nanoGPT/}, a medium-sized GPT \cite{gpt2} that tries to be smaller, cleaner, and more interpretable. With only a 6-layer Transformer \cite{vaswani2017attention} with 6 heads, it is very easy to hack it to our need, run on a single GPU or even CPU, and train it from scratch to fit our prompt-completion pairs data. As a GPT model, it decodes texts auto-regressively as Equation \ref{eq:LM} shown.

\subsection{Training}\label{subsec:training}
We train the NanoGPT with the prompt-completion pairs. Specifically, we combine all prompt-completion pairs into one corpus. The corpus will be sliced into contexts with a size of 256 characters, and each context will be tokenized into a feature with 384 channels. Then the NanoGPT takes the tokens as input to perform the next characters in a teacher-forcing paradigm \cite{sutskever2014sequence}. The cross-entropy loss will be calculated to update the model parameters: 

\begin{equation}
    \mathcal{L} = -\frac{1}{N} \sum_{i=1}^{N} \log p(c_i \mid c_{<i}),
\end{equation}where $N$ is the number of characters in the input sequence, $c_i$ is the $i$-th character, $c_{<i}$ is the sequence of previous characters, and $p(c_i | c_{<i})$ is the probability of predicting the next character $c_i$ given the previous characters $c_{<i}$. The cross-entropy loss $\mathcal{L}$ measures the difference between the predicted probability distribution and the true distribution of the next character.

\subsection{Inference}\label{subsec:inference}
In the inference phase, we can only use the music to get the prompt tokens. The trained NanoGPT will take the prompt tokens as input to predict the completion, i.e. the RGB-X tuples. With the RGB-X tuples predicted, a frame can be generated by a post-process script, since a frame can be completely determined by a set of RGB-X tuples. Finally, the frames will form a Launchpad-playing video.

\section{Experimental Results}\label{sec:experiment}

\subsection{Dataset}\label{subsec:dataset}
We have collected 16 Launchpad-playing videos from the Internet, with a total duration of 3312 seconds. The frame rates of these videos are 25 Hz. Therefore, we have about 82800 frames in total. To ensure a uniform form, we concatenate all videos together to get a long video and crop each frame to 128$\times$128 images. We cut the cropped long video into slices with a duration of about 10 seconds (250 video frames). Finally, the audio (music) of each video is extracted separately. These video frames and audio of music can produce about 82800 prompt-completion pairs, where 90\% are for training and 10\% are for validation.
\subsection{Metric}\label{subsec:metric}

We evaluate the performance of music visualization design on Launchpad with Fréchet Video Distance (FVD) \cite{unterthiner2019fvd} metric. FVD is a metric used to measure the similarity between the feature representations of real and generated videos. It is an extension of the Fréchet Inception Distance (FID) \cite{heusel2017gans} metric, which is used to evaluate the quality of generated images. FVD uses a pre-trained neural network as a feature extractor to compute the distance between two sets of video features. The formula for FVD can be expressed as:

\begin{equation}
    FVD(X,Y) = ||\mu_X - \mu_Y||^2 + cov(X,Y),
\end{equation}
\begin{equation}
    cov(X,Y)=Tr(\Sigma_X + \Sigma_Y - 2{(\Sigma_X\Sigma_Y)}^{\frac{1}{2}}),
\end{equation}
where $X$ and $Y$ are the sets of feature representations of real and generated videos, $\mu_X$ and $\mu_Y$ are the mean feature vectors of $X$ and $Y$, and $\Sigma_X$ and $\Sigma_Y$ are the covariance matrices of $X$ and $Y$. The FVD score is calculated by comparing the distance between the mean feature vectors and the trace of the covariance matrices of the two videos. We represent the FVD score as FID$\downarrow$, which means the smaller the score, the better.

\subsection{Quantitative Experiment}\label{subsec:quantitative_experiment}
We use two random methods, \textit{Random-RGB} and \textit{Random-RGBX} as the baselines to compare with our proposed LaunchpadGPT:
\begin{itemize}
    \item \textbf{Random-RGB.} The colors of the three channels of all 64 buttons are determined randomly.
    \item  \textbf{Random-RGBX.} The buttons that will be activated are determined randomly at first. Then the RGB color of these selected buttons will be decided randomly.
\end{itemize}

The experiment results are shown in Table \ref{tab:fvd}. It shows that the Launchpad-playing videos produced by our proposed LaunhpadGPT have the lowest FVD$\downarrow$ among all the methods, meaning LaunchpadGPT can generate Launchpad-playing videos that are more similar to the videos produced manually by humans.

\begin{table}[h]
\begin{center}
\begin{tabular}{|l|l|}
\hline
Method                       & FVD$\downarrow$            \\ \hline
Random-RGB                   & 350.63         \\ \hline
Random-RGBX                  & 147.64         \\ \hline
\textbf{LaunchpadGPT (ours)} & \textbf{75.22} \\ \hline
\end{tabular}
\end{center}
 \caption{The quantitative experiments. The result shows the Launchpad-playing videos produced by proposed LaunhpadGPT have the lowest FVD$\downarrow$ among all the methods.}
 \label{tab:fvd}
\end{table}

\subsection{Result Visualization}\label{subsec:visualization_result}
\begin{figure}[h]
  \centering
  \includegraphics[width=\columnwidth]{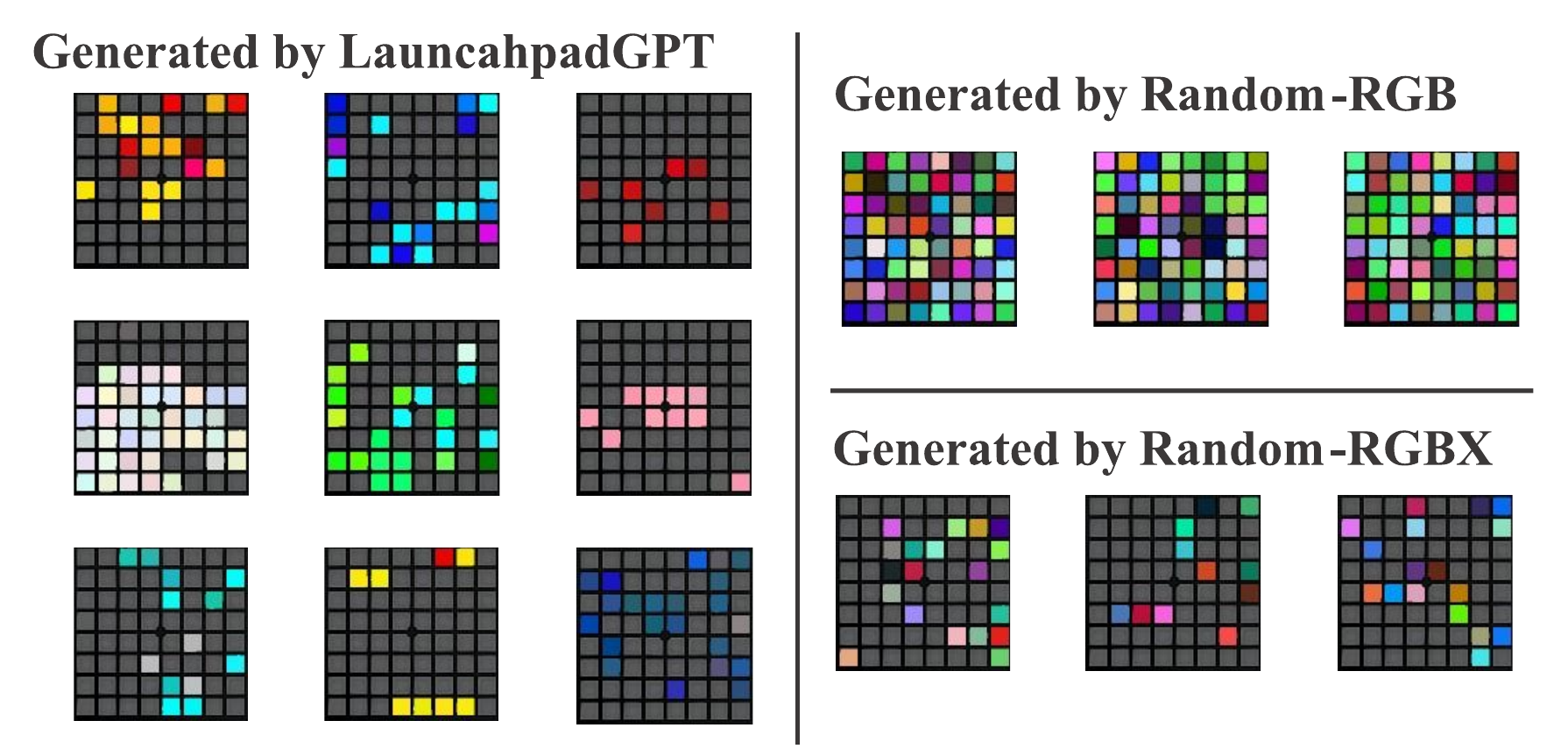}
  \caption{The outputs generated by LaunchpadGPT, Random-RGB, Random-RGBX respectively.}
  \label{fig:output}
\end{figure}

The language model output consists of RGB-X tuples and then be visualized as a colored Launchpad keyboard. 
As Figure \ref{fig:output} shows, we compare the output results from three different approaches: LaunchpadGPT, Random-RGB, and Random-RGBX. The observed results of outputs are illustrated as follows.

\begin{itemize}
% The Random-RGB approach
    \item \textbf{The outputs of Random-RGB} presents a chaotic and disorganized color effect on the Launchpad keyboard. As the colors are randomly generated, there is no discernible pattern or coherence in the color scheme.
    \item \textbf{The outputs of Random-RGBX} are random colors applied to illuminate specific button coordinates on the Launchpad keyboard. This results in a scattered pattern of randomly lit buttons, without any meaningful relationship between the colors and the button coordinates.
    % The output generated by LaunchpadGPT
    \item \textbf{The outputs of LaunchpadGPT} presents similarities in color tones. The RGB colors tend to belong to the same color spectrum or display slight variations. This indicates that the language model has learned to generate RGB colors that are aesthetically coherent and reveal a level of color similarity.
\end{itemize}

The results show that LaunchpadGPT performs better than the random methods. 
This finding demonstrates the model's ability to capture and reflect the similarity in color tones, contributing to the generation of visually coherent RGB-X tuples. However, while LaunchpadGPT captures color similarities to some extent, it falls short in learning more structured patterns. We assume that the limitation lies in the representation of the RGB-X tuples, as the X component does not effectively convey the spatial relationships between buttons.

\begin{figure}[h]
\centering
\includegraphics[width=1.0\columnwidth]{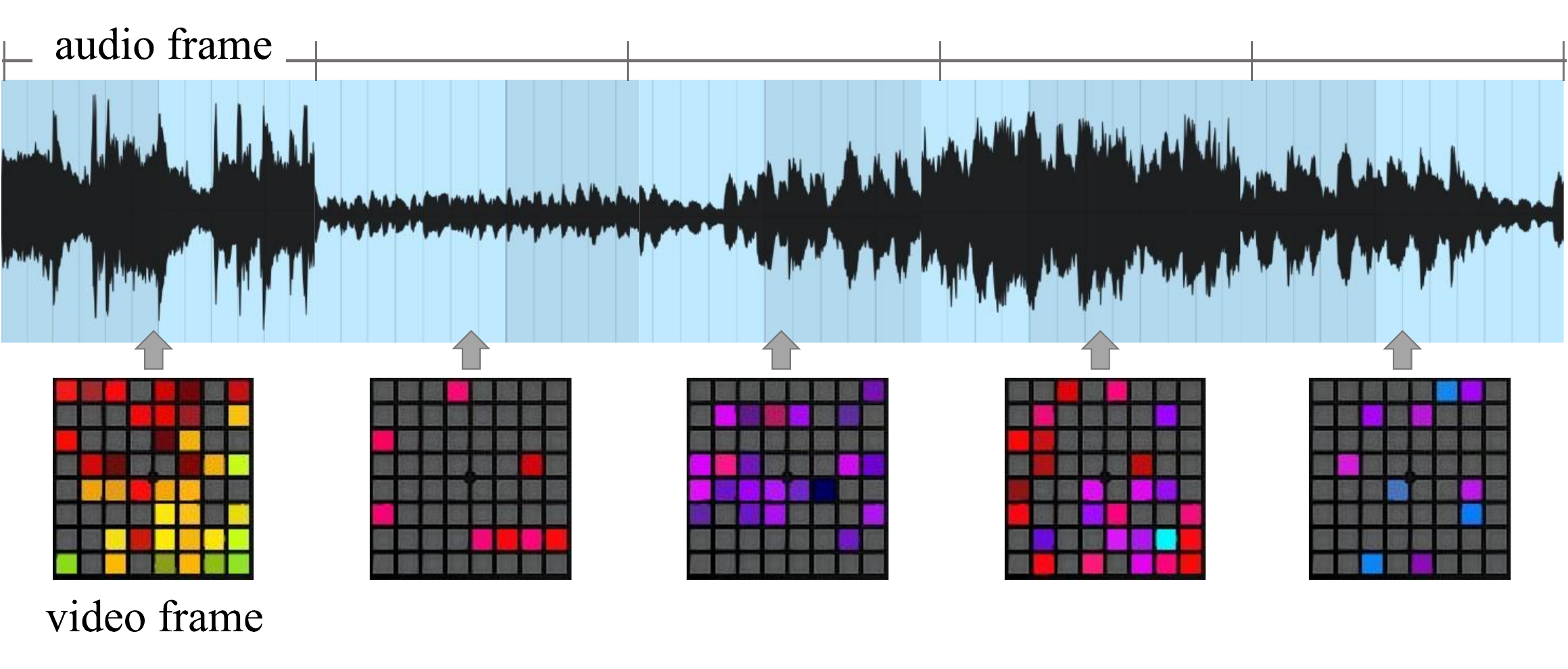}
\caption{Results visualization aligning audio frames with corresponding generated video frames}
\label{fig:bo}
\end{figure}

In addition, we also discovered a positive correlation between the pattern activity of generated frames and the amplitude of the music's audio frames, which is shown in Figure \ref{fig:bo}. As the amplitude of the audio frames increases, the corresponding generated video frames exhibit higher pattern activity level. Conversely, when the amplitude is lower, the pattern activity level decreases.

% TODO:更多定性分析结论待补充
The revealed ability of LaunchpadGPT opens up possibilities for task transfer in various music-related domains. The envisioned applications include designing music game charts, LED screen designs for music performance venues, and lighting schemes for music dance floors. By leveraging the model's ability to understand music-related patterns, designers, game developers, and lighting engineers can benefit from its generative capabilities to enhance the visual experience in music-related contexts.

% In summary, the qualitative analysis of the language model's output data highlights a consistent tendency towards color similarity or minor variations within a given color palette. This finding showcases the model's competence in understanding and generating RGB colors that adhere to the concept of color similarity. However, to ascertain the model's overall applicability and reliability in generating visually synchronized Launchpad keyboard representations, further investigation and validation are warranted.

% \pagebreak

\section{Conclusions}
We proposed LaunchpadGPT, a language model based on GPT, to generate music visualization on Launchpad with given music, assisting music designers in building Launchpad projects. LaunchpadGPT learns from Launchpad-playing videos to automatically generate synchronized light design schemes that enhance the visual experience. It simplifies the process of light design and empowers designers to explore different lighting schemes effortlessly, while also producing visually captivating playing videos showcasing the synchronized interaction between Launchpad, music, and dynamic lighting effects. The revealed potential of LaunchpadGPT in music visualization design can lead to a broader range of applications within the field of music visualization.

\begin{acknowledgments}
This work was supported by the National Key R\&D Program of China (Grant NO. 2022YFF1202903) and the National Natural Science Foundation of China (Grant NO. 62122035 and 61972188). 
\end{acknowledgments} 

%%%%%%%%%%%%%%%%%%%%%%%%%%%%%%%%%%%%%%%%%%%%%%%%%%%%%%%%%%%%%%%%%%%%%%%%%%%%%
%bibliography here
\bibliography{icmc2023template}

\end{document}